\documentclass[%
 reprint,
 prl,
superscriptaddress,
 amsmath,amssymb,
 aps,
]{revtex4-1}

\usepackage{comment}
\usepackage{dsfont}
\usepackage{graphicx}
\usepackage{dcolumn}
\usepackage{bm}
\usepackage{color}
\usepackage{epsfig}
\usepackage[colorlinks=true, linkcolor=blue, citecolor=blue, urlcolor=blue]{hyperref}
\usepackage{natbib}
\usepackage{float}
\usepackage{footmisc}
\usepackage{slashed}
\usepackage[dvipsnames]{xcolor}
\usepackage{braket}
\usepackage{soul}

\definecolor{lightred}{rgb}{1,0.5,0.5}
\definecolor{lightgreen}{rgb}{0.5,1,0.5}
\definecolor{lightblue}{rgb}{0.5,0.5,1}
\definecolor{lightcyan}{rgb}{0.5,0.75,0.75}
\definecolor{lightmagenta}{rgb}{0.75,0.5,0.75}
\definecolor{customgreen}{rgb}{0.494,1,0.502}

\newcommand{\meV}{\mathinner{\mathrm{meV}}}

\def\bea{\begin{eqnarray}}   \def\eea{\end{eqnarray}}
\def\mDM{m_{\rm DM}}

\usepackage{tikz,xcolor,hyperref}
\definecolor{lime}{HTML}{A6CE39}
\DeclareRobustCommand{\orcidicon}{%
	\begin{tikzpicture}
	\draw[lime, fill=lime] (0,0) 
	circle [radius=0.16] 
	node[white] {{\fontfamily{qag}\selectfont \tiny ID}};	\draw[white, fill=white] (-0.0625,0.095) 
	circle [radius=0.007];	\end{tikzpicture}
	\hspace{-2mm}}
\foreach \x in {A,...,Z}{%
	\expandafter\xdef\csname orcid\x\endcsname{\noexpand\href{https://orcid.org/\csname orcidauthor\x\endcsname}{\noexpand\orcidicon}}
}

\def\UD{\small{Department of Physics and Astronomy, University of Delaware, Newark, Delaware 19716, USA}}

\def\UMD{\small{Maryland Center for Fundamental Physics, University of Maryland, College Park, MD 20742, USA}}

\def\UMDphysics{\small{Department of Physics, University of Maryland, College Park, Maryland 20742, USA}}

\def\JHU{\small{Department of Physics \& Astronomy, The Johns Hopkins University, Baltimore, MD 21218, USA}}

\begin{document}
\title{Rydberg Single Photon Detection for Probing 0.1–10 meV Dark Matter with BREAD}

\author{Abhishek~Banerjee\orcidA{}\,}
\email{abanerj4@umd.edu}
\affiliation{\UMD}
\author{Reza~Ebadi\orcidB{}\,}
\email{ebadi@umd.edu}
\affiliation{\JHU}
\affiliation{\UD}
\affiliation{\UMDphysics}
\author{Surjeet~Rajendran\orcidC{}\,}
\email{srajend4@jhu.edu}
\affiliation{\JHU}

\begin{abstract}
    We introduce a Rydberg–based single photon detector (SPD) for probing dark matter in the 0.1--10\,meV mass range (20\,GHz--2\,THz). The Rydberg SPD absorbs photons produced and focused by the BREAD dish antenna and trades them for free, detectable electrons. At the lower end of the mass range, photons drive Rydberg–Rydberg transitions, which are read out via state-selective ionization. At higher masses, they directly ionize the Rydberg atoms.
\end{abstract}

\maketitle

Light bosonic dark matter candidates (e.g. axion, CP-even scalars, and dark photon) are well described as coherent classical fields~\cite{Kolb:1990vq, Cheong:2024ose}. 
Through their coupling to electromagnetism, a small fraction of the background dark matter (DM) converts into coherent photons. The non-relativistic nature of DM implies that the photon frequencies are narrowly distributed around the dark matter mass $\omega \simeq m_{\rm DM}$. Numerous existing and proposed experiments are designed to search for such DM–induced ac electromagnetic signal \cite{Adams:2022pbo, Sushkov:2023fjw}.

\begin{figure}[ht]
    \centering
    \includegraphics[width=\linewidth]{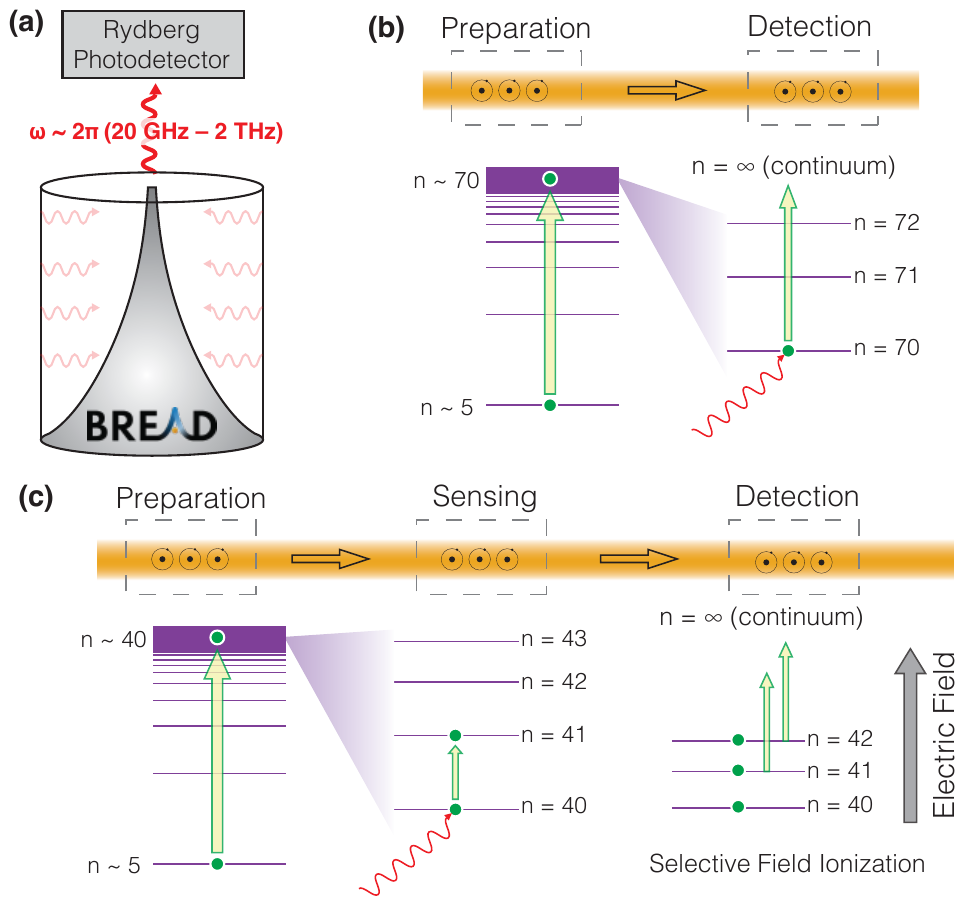}
    \caption{Schematic of the Rydberg SPD setup. {\bf (a)} BREAD uses a parabolic reflector to focus photons emitted perpendicular to the barrel surface onto a focal point, increasing the photon intensity at the focus. {\bf (b)} A Rydberg beam is prepared by exciting electrons to a high principal quantum number. When the photon energy exceeds the ionization threshold $\omega > I_n$ the photons directly ionize the Rydberg atoms, producing detectable free electrons. {\bf (c)} The target photon energy matches the transition energy between Rydberg states, $\omega = \omega_n$. During the sensing stage, the Rydberg beam is exposed to photons from the BREAD setup. Downstream, the Rydberg state is read out using selective field ionization (SFI). In SFI, an external electric field is scanned across the ionization threshold of the relevant states, and the resulting ionization is detected.}
    \label{fig:schematics}
\end{figure}

The Broadband Reflector Experiment for Axion Detection (BREAD) employs a dish antenna with a parabolic reflector to focus photons generated by DM interactions onto a focal point \cite{BREAD:2021tpx}. BREAD is a modular broadband signal-enhancement platform that must be supplemented with a high-precision readout module. Developing efficient detectors continues to be an active area of research. A key focus is the development of single photon detectors (SPDs), as photon-counting is expected to achieve higher sensitivity in the low-signal regime compared to bolometric detection \cite{Sushkov:2023fjw}.

SPDs at optical and infrared frequencies are mature technologies \cite{hadfield2009single}. However, efficient single photon detection in the THz range remains challenging \cite{todorov2024thz}. Photons in this frequency range carry energies on the meV scale, far lower than the eV-scale energies of optical photons. Thus, well-established optical SPDs lack sensitivity in this range. Addressing this challenge requires development of new detection technologies. The search for meV-scale dark matter has recently driven further development of THz SPDs \cite{Fan:2024mhm}.

Graham {\it et al.} proposed a Rydberg-based SPD for the HAYSTAC experiment to probe DM in the 40--200\,${\rm \mu eV}$ mass range (10--50\,GHz) \cite{Graham:2023sow}. In this setup, DM–induced photons drive transitions between Rydberg states, and the resulting state population is read out via selective field ionization (SFI). Inspired by this approach, we introduce a Rydberg-atom–based SPD operating in the 20\,GHz--2\,THz range as a complementary component of the BREAD experiment. Rydberg electronic energy levels span this frequency range, and their large dipole coupling enhances photon absorption efficiency. A photon absorbed by a Rydberg atom can either directly ionize it or excite it to a higher Rydberg state (detectable via state-selective field ionization); see Fig.\,\ref{fig:schematics}. 

One of our proposed detection schemes (Rydberg transitions induced by photons followed by state readout via SFI) is similar to that in Ref.\,\cite{Graham:2023sow}. However, our approach differs in several key aspects. First, the direct ionization scheme we introduce is entirely new and enables sensitivity to DM masses substantially higher than those accessible through transition-based methods. Second, while Ref.\,\cite{Graham:2023sow} considers a microwave cavity setup with separate cavities for signal buildup and detection, our design employs the broadband reflector configuration of BREAD. This setup is considerably simpler, as it removes the need for frequency scanning and allows searches at higher frequencies using Rydberg ionization, since the signal strength is no longer limited by the cavity size.

{\textbf{Photons from dark matter.---}}
We focus on two specific DM candidates: the axion and the dark photon. The relevant electromagnetic interaction Lagrangian takes the form $\mathcal{L}\supset -g_{a\gamma\gamma} a F\tilde F/4 = g_{a\gamma\gamma} a\mathbf{E}\cdot\mathbf{B}$ for the axion, and $\mathcal{L}\supset \varepsilon F F'/2$ for the dark photon. Here, $a$ is the pseudoscalar axion field, $g_{a\gamma\gamma}$ the axion–photon coupling, $\varepsilon$ the dimensionless dark photon–photon mixing parameter, $F$ the electromagnetic field strength, and $F'$ the dark photon field strength. When a large bias magnetic field $\mathbf{B}_0$ is applied, ambient axion DM produces an oscillating electric field with amplitude $E_{\rm a} \sim g_{a\gamma\gamma} B_0 \sqrt{\rho_{\rm DM}} / m_{\rm DM}$. For the dark photon, the mixing automatically produces a similar oscillating electric field $E_{A'} \sim \varepsilon \sqrt{\rho_{\rm DM}}$ where $\rho_{\rm DM}$ denotes the local DM energy density.

Coherent detection of the ac electric field is linearly sensitive to the couplings and has been proposed for DM detection using electro-optics \cite{Ebadi:2023gne}. In the mass range of interest for this work, however, the photon production by BREAD occurs incoherently. For $m_{\rm DM} \gtrsim 0.1\,{\rm meV}$, the DM de Broglie wavelength is smaller than the size of the BREAD setup \cite{BREAD:2021tpx}. As a result, photons emitted from different parts of the barrel surface are out of phase, producing an incoherent signal. We will therefore focus on single photon detection, which has quadratic sensitivity to the couplings. While the standard quantum limit sets the fundamental limit for coherent detection, the dark count rate (DCR) is the primary figure of merit for SPDs.

The BREAD setup will have a dish area of $A_{\rm dish}\sim10\,{\rm m}^2$. The focus area $A_{\rm focus}$ is smeared out compared to an ideal point due to the momentum spread of the emitted photons, inherited from the momentum spread of the DM. With a typical momentum spread of $\Delta p_{\rm DM}/m_{\rm DM}\sim 10^{-3}$ and a physical BREAD setup size of about a meter, the resulting focus area is estimated as $A_{\rm focus}\sim \pi \,{\rm mm}^2$ \cite{BREAD:2021tpx}. Note that the photon field amplitude at the focal point is enhanced by a factor of $\sqrt{A_{\rm dish}/A_{\rm focus}}\sim\mathcal{O}(10^3)$ relative to its value away from the focus.

We compute the signal photon rate by estimating the emitted power $P_{\rm DM}=E_{\rm DM}^2A_{\rm dish}/2$ where $A_{\rm dish}$ is the effective emission area. The corresponding photon rate is $R_{\rm DM}=P_{\rm DM}/m_{\rm DM}$. Therefore, the expected signal photon rates are $R_a \simeq g_{a\gamma\gamma}^2 B_0^2 \rho_{\rm DM} A_{\rm dish}/2m_a^3$ and $R_{A'} \simeq \varepsilon^2 \rho_{\rm DM} A_{\rm dish}/2m_{A'}$. For fixed external parameters $B_0 = 10~\mathrm{T}$, $\rho_\mathrm{DM}=0.45~\mathrm{GeV/cm^3}$, and $A_\mathrm{dish}=10~\mathrm{m^2}$, these expressions provide numerical estimates of the photon rates as a function of couplings and masses \cite{BREAD:2021tpx}:
\begin{align}
        R_{a} &\simeq 0.55~\mathrm{Hz} \bigg(\frac{g_{a\gamma\gamma}}{10^{-11}~\mathrm{GeV^{-1}}}\bigg)^2 \bigg(\frac{m_a}{\mathrm{meV}}\bigg)^{-3}, \nonumber \\
        R_{A'} &\simeq 0.14~\mathrm{Hz} \bigg(\frac{\varepsilon}{10^{-14}}\bigg)^2 \bigg(\frac{m_{A'}}{\meV}\bigg)^{-1}. \nonumber
\end{align}

DM–photon conversion can be further enhanced by introducing a stack of $N_l$ dielectric layers, which effectively increases the number of conversion surfaces \cite{Jaeckel:2013eha,Caldwell:2016dcw,Baryakhtar:2018doz}. In the BREAD setup, this approach can enhance the photon rate by a factor of $N_l^2/2$, where the optimal number of layers is $N_l \simeq 80 \sqrt{m_{\rm DM}/{\rm meV}}$ \cite{Fan:2024mhm}. Thus, the signal photon rate will be enhanced by a factor of $10^3 (m_{\rm DM}/{\rm meV})$. At the same time, the required semi-resonant buildup across the layers limits the emitted radiation to $Q \sim N_l/N_s$, where $N_s$ is the number of stacks with different spacings. Ref.\,\cite{Fan:2024mhm} identified that for the optimal case $Q\sim17$, well below the intrinsic quality factor of the DM wave $Q_{\rm DM} \sim 10^6$.

\renewcommand{\arraystretch}{1.3} 
\begin{table}[t]
\centering
\caption{Scaling behavior of Rydberg quantities.}
\begin{tabular}{l|c|c}
\hline
\hline
Quantity & Symbol & Scaling \\
\hline
\hline
Typical orbital size & $\langle r \rangle_{n}$ & $\sim n^2/m_e\alpha$ \\
Binding energy &  $I_{n}$ & $\sim m_e\alpha^2/2n^2$ \\
Transition frequency & $\omega_{n}$ & $\sim m_e\alpha^2/n^3$ \\
Radiative lifetime & $\tau_{n}$ & $ \sim \begin{cases} 
100\,(n/40)^3\,{\rm \mu s}, & l = 1 \\ 
10\,(n/40)^5\,{\rm m s}, & l = n-1 
\end{cases}$ \\
dc ionizing field &  $E_{\rm ion}$ & $\sim 30 \, n^{-4}\,{\rm GV/m}$ \\
\hline
\end{tabular}
\label{tab:rydberg_scaling}
\end{table}

{\textbf{Rydberg SPD.---}} A Rydberg atom has an electron excited to a very high principal quantum number $n\gg1$~\cite{Gallagher1994}. 
Rydberg atoms exhibit exaggerated physical properties; see Table~\ref{tab:rydberg_scaling}. A defining feature is their large orbital size, which scales as $\sim n^2 a_0$, where $a_0$ is the Bohr radius~\footnote{For Rydberg atoms, the deviation from the pure Coulomb potential due to the presence of core electrons is incorporated through the quantum defect, $\delta_\ell$, which depends on $\ell$. 
Thus an effective description is obtained by replacing $n\to n^*=n-\delta_l, \ell\to\ell^*=l-\delta_l+I(l)$ where $I(\ell)$ is some integer~\cite{PhysRevA.32.3243}. 
This is a small correction to the hydrogenic values and do not alter the physics discussed here.}. This leads to strong dipole couplings, both between Rydberg atoms and with photons. We exploit the strong photon–Rydberg interaction to detect DM–induced photons, for the same reason it has been used in similar applications \cite{Jau_2020,Meyer_2020,li2024room,Nill_2024_avalanche,liu_2023_efield}.

We consider two Rydberg-based sensing modalities: Rydberg–Rydberg transitions and Rydberg–continuum ionization (see Fig.\,\ref{fig:schematics}). Lower-energy photons (0.1--1\,meV) can drive Rydberg–Rydberg transitions, while higher-energy photons (1--10\,meV) can directly ionize electrons from the Rydberg state. In the Rydberg–Rydberg transition case, the final state can be read out via state-selective field ionization (SFI) \cite{hollenstein2001selective,Tada_2002,GURTLER2004315,Don_Fahey_SFI_2015,Noel_SFI_2017,Alonso_2018_sfi,Gregoric_2018_sfi}. In SFI, an external electric field is ramped up until the atom ionizes, and the free electron is detected. Different Rydberg levels ionize at distinct field strengths, providing a correspondence between the applied field and the state. In the direct ionization case, the electron is detected immediately.

Rydberg states with lifetimes $\tau_n \gtrsim 10\,\mu{\rm s}$ exhibit narrow transitions with quality factor $Q_{\rm Ryd}=\omega_n/(2\pi/\tau_n)\gtrsim10^7$. Because this exceeds the DM-induced photon quality factor $Q_{\rm DM}\sim10^6$ for BREAD and $\sim 20$ for BREAD with dielectric layers, the effective search bin width is determined by the photon linewidth rather than the transition linewidth. In contrast, ionization occurs whenever the DM-induced photon energy exceeds the ionization threshold $I_n$ and lies within the effective ionization range. The ionization search is therefore broadband, determined by the readout scheme and independent of the signal linewidth.

The transition energy between neighboring Rydberg states is $\omega_n\simeq27.2\,{\rm eV}/n^3$, corresponding to $n \in (30,65)$ for the 0.1--1\,meV DM mass range. The binding energy is $I_n\simeq13.6\,{\rm eV}/n^2$, corresponding to $n \in (37,120)$ for the 1--10\,meV range. Note that not all of these $n$ values are required to cover the DM mass range of interest; we discuss the full mass coverage strategy later.

The rates of transition and ionization relative to the photon production rate serve as a figure of merit for how efficiently the BREAD-produced photons are collected. We compute these rates for low-$l$ states using Fermi’s golden rule; see the Supplemental Material for details. For a single Rydberg atom, the transition rate between a Rydberg state $n$ and a nearby state $n+i$, where $i \ll n$, is $\gamma_{n\rightarrow n+i} \simeq (2\pi\alpha/3) E_{\rm DM}^2 \, n^4 a_0^2\, {\rm min}[\tau_{\rm DM},\tau_{n},\tau_{\rm sens.}]$. Here, $\tau_{\rm DM},\tau_{n}$, and $\tau_{\rm sens.}$ are the DM-induced photon coherence time, the Rydberg state lifetime, and the time each atom spends in the sensing (focal) region, respectively. The Rydberg ionization rate is better understood away from the ionization threshold \cite{BetheSalpeter1957}, since Coulomb corrections become significant near the threshold. We adopt a conservative estimate $\gamma_{n \rightarrow {\rm cont.}} \simeq (1024\pi/3)E_{\rm DM}^2 n^4 a^3_0\,(I_n/m_{\rm DM})^6 \left(m_{\rm DM}/I_n-1\right)^{3/2}$. The ionization rate has broad support in DM mass, with a power-law fall-off. The maximum value is $\gamma_{n \rightarrow {\rm cont.}}^{\rm max} \simeq 36 E_{\rm DM}^2 n^4 a^3_0$. The estimated rates are consistent in order of magnitude with values reported in the literature, and show only mild $l$-dependence for low-$l$, high-$n$ states~\cite{Ovsiannikov2011}. 
The transition rates between nearby $n$ levels also exhibit only subleading dependence on $l$. This is unlike the radiative lifetime that has a strong $l$-dependence. The reason is that the dominant radiative decay channels for low-$l$ states correspond to transitions to much lower $n$ states (for which a larger phase space is available), and as a result, the dipole matrix element acquires a distinct overall $n$-scaling behavior.

$E_{\rm DM}$ is the photon field amplitude at the focusing area $E_{\rm DM}^2=2 R_{\rm DM} m_{\rm DM}/A_{\rm focus}$. 
Key advantage of a focusing setup such as BREAD is manifest in these rates: the photon field amplitude at the detector site is enhanced. Since $R_{\rm DM}\propto A_{\rm dish}$, the rate enhancement scales as $A_{\rm dish}/A_{\rm focus}\sim 3\times10^6$ for typical BREAD parameters.

In our proposed method, a large number of Rydberg atoms simultaneously interact with photons in the focused area. Assuming a Rydberg beam with flux $\Phi_{\rm beam}\sim 10^{15}\,{\rm cm^{-2} s^{-1}}$ and velocity $v_{\rm beam}\sim{\rm km/s}$, the total number of Rydberg atoms present at any instant in the sensing volume is $N_{\rm Ryd} \simeq \Phi_{\rm beam} A_{\rm focus} \tau_{\rm sens.}\sim 3\times10^7$, where $\tau_{\rm sens.}=\sqrt{A_{\rm focus}/\pi}/v_{\rm beam}\simeq{\rm \mu s}$. 

The total photon absorption rate is given by the single-atom rate multiplied by the number of Rydberg atoms $\Gamma=N_{\rm Ryd}\gamma$. We define efficiency of photon absorption as $\eta = \Gamma/R_{\rm DM}$. For BREAD ($Q_{\rm DM}\sim10^6$), the limiting timescale is $\tau_{\rm sens.}$: ${\rm min}[\tau_{\rm DM},\tau_{n},\tau_{\rm sens.}] = \tau_{\rm sens.}$. For BREAD with the dielectric layers ($Q_{\rm DM}\sim 20$), the limiting timescale is instead the DM-induced photon coherence time: ${\rm min}[\tau_{\rm DM},\tau_{n},\tau_{\rm sens.}] = \tau_{\rm DM}$. The corresponding {\it nominal} efficiencies are:
\begin{align}
        \eta_{n\rightarrow n+i}^{\rm BREAD} &\simeq 1.5\times 10^3 \bigg(\frac{m_{\rm DM}}{0.3\,{\rm meV}}\bigg)^{-1/3} \bigg(\frac{\Phi_{\rm beam}}{10^{15}\,{\rm cm^{-2} s^{-1}}}\bigg), \nonumber\\
        \eta_{n\rightarrow n+i}^{\rm Dielectric} &\simeq 0.3 \bigg(\frac{Q_{\rm DM}}{17}\bigg) \bigg(\frac{m_{\rm DM}}{0.3\,{\rm meV}}\bigg)^{-4/3}  \bigg(\frac{\Phi_{\rm beam}}{10^{15}\,{\rm cm^{-2} s^{-1}}}\bigg), \nonumber\\
        \eta_{n\rightarrow {\rm cont.}}^{\rm max} &\simeq 3\times10^{-5} \bigg(\frac{m_{\rm DM}}{3\,{\rm meV}}\bigg)^{-1} \bigg(\frac{\Phi_{\rm beam}}{10^{15}\,{\rm cm^{-2} s^{-1}}}\bigg), \nonumber
\end{align}
where we have used the scaling relations $m_{\rm DM}\propto n^{-3}$ for transition and $m_{\rm DM}\propto n^{-2}$ for ionization. 
The efficiency cannot exceed unity, so the effective efficiency is simply taken to be the smaller of the nominal value and one.

To summarize, in the high-Q resonant modality of BREAD, the Rydberg SPD can detect essentially all of the photons produced. In the lower-Q BREAD setup with dielectric layers, the Rydberg SPD remains highly efficient but may not absorb every photon. However, since this search is broadband compared to the high-Q modality, the overall sensitivity improves due to relatively longer averaging time available per experiment. In the ionization modality, the Rydberg SPD collects only a small fraction of the photons, but since the search is even broader in bin size, it still provides competitive sensitivity across the higher-mass parameter space.

\begin{figure}[t]
    \centering
    \includegraphics[width=\linewidth]{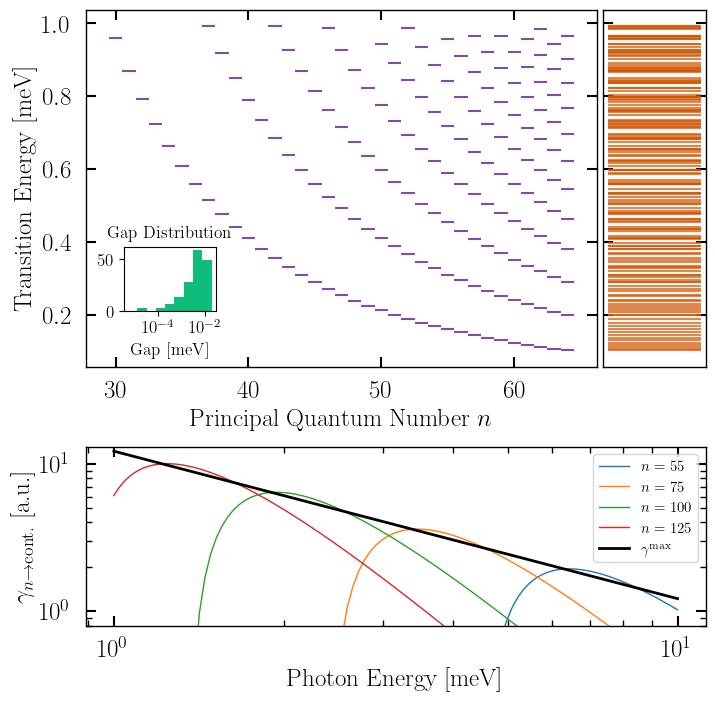}
    \caption{Dark matter mass range accessible using Rydberg states with principal quantum number $n$. \textit{{Top:}} Rydberg transition coverage (without scanning) for $n=30, 31, \cdots, 65$. We include transitions of $n\rightarrow n+i$ within the 0.1-1\,meV range. The inset shows the distribution of mass gaps across all available transitions. The largest uncovered gap in mass coverage is 0.02\,meV, which can be bridged using a perturbative scanning scheme in about $0.02\,{\rm meV}\times Q_{\rm DM}/m_{\rm DM}\sim4\times10^4$ steps for $Q_{\rm DM}\sim10^6$. \textit{{ Bottom:}} Ionization is efficient for photons with frequencies above the ionization threshold, with a power-law fall-off at higher frequencies. This enables broadband sensitivity using only a small number of atoms. Using four different Rydberg states allows coverage of the 0.1--1 meV DM mass range.}
    \label{fig:scanning}
\end{figure}

{\textbf{Experimental feasibility.---}} Rydberg states with $n$ up to 100 can be readily prepared \cite{Jau_2020}. Toward the upper end of this range, however, Rydberg states become increasingly fragile to stray electric fields. The ionizing dc electric field is $\sim 3\,\mathrm{V/cm}$ for $n \sim 100$, which can be shielded against. At lower $n$, the ionizing electric field amplitude increases rapidly as $n^{-4}$.

Rydberg states are relatively long-lived \cite{Branden_2009,Mack_2015,Holzl_2024}. Their main decay channels are spontaneous emission and blackbody-radiation (BBR)–induced transitions. The BBR rate is suppressed relative to spontaneous decay by the Boltzmann factor $1/(e^{\omega/k_B T}-1)$, which is $\lesssim 10^{-2}$ at $T \sim 1\,{\rm K}$ in the frequency range of interest. At such low temperatures, the lifetime is limited by spontaneous decay. For circular states ($l=n-1$) it is expected to be $10 (n/40)^5\,{\rm ms}$, while for $P$ states ($l=1$) it is shorter, $100 (n/40)^3\,\mu{\rm s}$ \cite{Sibalic:2017agz}. The reduced lifetime of low-angular-momentum states arises from their larger overlap with low-$n$ core states.

In both detection modalities (direct ionization and transition followed by SFI) the freed electron must be collected. A channel electron multiplier (CEM) can be used for this purpose \cite{Graham:2023sow}. CEMs are widely used in similar contexts, for example in photoionization detection of a single atom \cite{Henkel_2010}. Imaging of electrons and ions produced by SFI has also been demonstrated \cite{Done_Fahey_dd_2015} using microchannel plates \cite{BIS}. Commercial devices \cite{Sjuts} achieve near-unity efficiency with dark count rate (DCR) of about $0.01\,{\rm Hz}$. Typical operating bias voltage for CEMs is $\gtrsim 100\,{\rm V}$. For the highest-$n$ states considered here, this corresponds to an electric field that can approach or exceed the ionization threshold. To avoid this, we propose using a low-voltage extraction electrode near the sensing region, followed by electrostatic steering of the electron toward the CEM. Local acceleration at the device’s operating voltage can then raise the electron to the detection threshold energy.

The Rydberg readout module operates at cryogenic temperatures to suppress BBR–induced backgrounds. On the lower end of the mass range of interest the required temperature to keep the BBR background rate below the electronic DCR of 0.01\,Hz is approximately 50\,mK~\cite{Graham:2023sow}. At higher masses, this requirement becomes less stringent, scaling as $500\,{\rm mK}\,(m_{\rm DM}/{\rm meV})$.

A Rydberg beam can be prepared from a high-intensity atomic beam with fluxes in the range $\Phi \sim 10^{11}$–$10^{16}\,{\rm cm^{-2}s^{-1}}$ \cite{larsen1974seeded,catani_2006,wei2022collimated}. Preparation can be achieved using two- or three-photon transitions \cite{Tate_2018,Done_Fahey_dd_2015} or stimulated Raman adiabatic passage (STIRAP) \cite{Cubel_2005}. We also note that Rydberg beams composed of low-$l$ states are the most straightforward to prepare, since with each absorbed photon only transitions with $\Delta l = \pm 1$ are allowed by the dipole selection rules. Our proposal therefore makes use of these readily accessible Rydberg states.

Given a Rydberg beam flux of $\Phi \sim 10^{15}\,{\rm cm^{-2}s^{-1}}$ and a beam velocity $v_{\rm Ryd} \sim {\rm km/s}$, the Rydberg atom number density is $n_{\rm Ryd} \sim 10^{10}\,{\rm cm^{-3}}$. This corresponds to an average spacing of about $5\,{\rm\mu m}$ between atoms, much larger than the size of the largest Rydberg state considered here ($n=100$) which is $10^4 a_0 \simeq 0.5\,{\rm\mu m}$. Therefore, the dipole–dipole interactions are suppressed \cite{Rydberg_review_2010}. Another potential factor that can contribute to efficiency reduction or backgrounds is atomic collisions within the Rydberg beam, which can be mitigated by engineering low-temperature beams. Ultimately, magneto-optical trap (MOT) cooling combined with conveyor-belt atomic transport featuring a continuously high reloading rate can be employed as the sensing beam in our proposal \cite{Wieman_2000,Aidelsburger_2022,Cornish_2024,Chiu:2025uis}.

The length of the Rydberg beam is $\tau_n v_{\rm Ryd} \gtrsim 10\,{\rm cm}$, assuming a lifetime $\tau_n \gtrsim 100\,{\rm \mu s}$. For both Rydberg preparation and SFI detection we assume operation times of about ${\rm \mu s}$, which are much shorter than $\tau_n$. Consequently, most of the beam length is available for sensing and transport between the different detection stages; see Fig.\,\ref{fig:schematics}. Minimizing the separation between detection stages reduces the dead time of each measurement. This would also suppress the lost Rydberg population fraction from stochastic decays $1 - e^{-3\,{\rm \mu s}/\tau_n}$ to the percent level or lower. This effect is negligible for our efficiency estimates. Also note that DM-induced transitions occur between nearby $n$ states, whereas radiative decays in low-$l$ states predominantly drive electrons to the lowest-$n$ levels. Therefore, we do not expect these decay processes to be a significant background for our $n \rightarrow n+i$ transition–based search.

As shown in Fig.\,\ref{fig:scanning}, the DM mass range 0.1-1\,meV can be covered using transitions with $n\in(30,65)$. The largest gap between adjacent transitions is $0.02\,{\rm meV} = 2\pi (4.8\,{\rm GHz})$. To bridge this gap, the transition frequencies can be shifted using external Zeeman or Stark effects. The Zeeman (magnetic) shift is, to leading order, independent of $n$: $\sim 2\pi\,{\rm MHz}(B_{\rm ext}/10\,{\rm G})$. In contrast, the Stark shift depends on $n$ quadratically: $\sim 2\pi(4\,{\rm GHz})(E_{\rm ext}/{\rm V/cm})(n/40)^2$. Therefore, modest magnetic fields (particularly at the lower end of this mass range) and electric fields at the V/cm level (at the upper end) are sufficient to fully cover the target mass range.

\begin{figure*}[t]
    \centering
    \includegraphics[width=\linewidth]{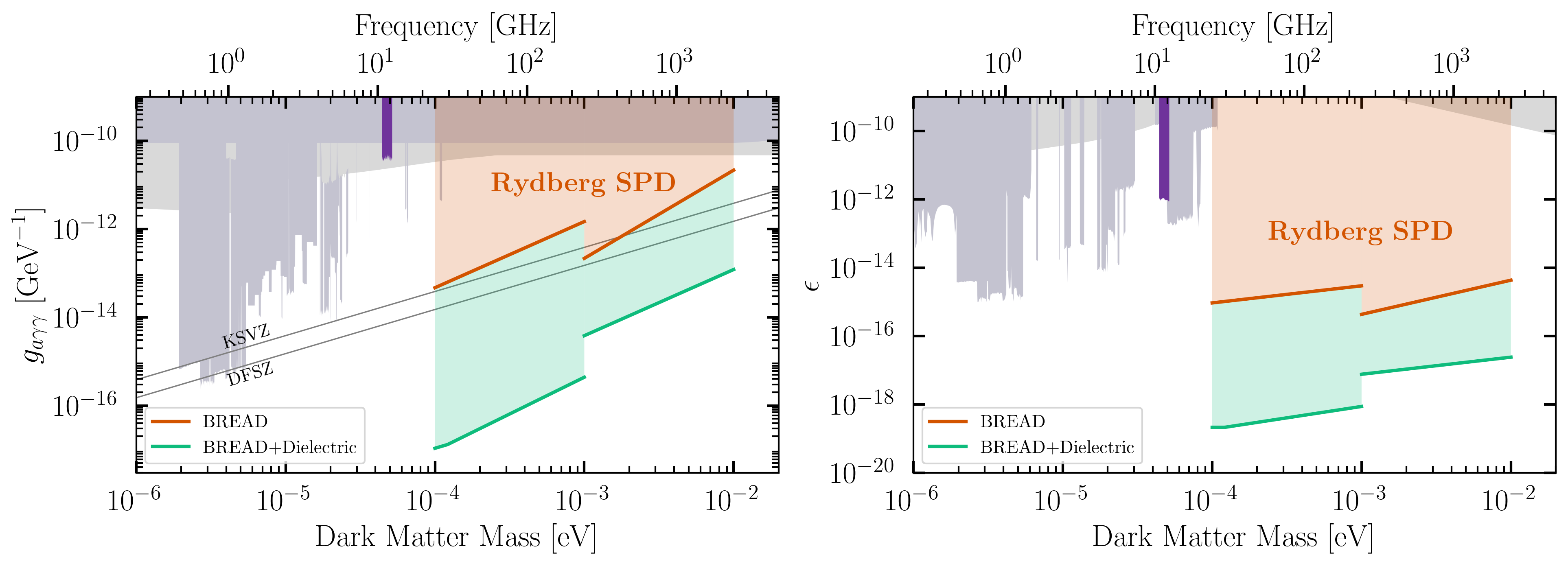}
    \caption{Projected Rydberg SPD sensitivities in the BREAD experiment for axion (left) and dark photon (right) dark matter. We assume SNR=5, DCR=0.01\,Hz, 1000\,days of measurement to cover the 0.1--1\,meV range, and an additional 1000\,days to probe the 1--10\,meV range. Orange lines correspond to the BREAD dish antenna setup, and green lines include additional dielectric layers that further enhance the photon production rate and also broaden the photon linewidth. Details are discussed in the main text. Existing constraints are shown in light gray (astrophysical and cosmological) and light purple (laboratory-based). Existing limits are adapted from Ref.\,\cite{AxionLimits} and references therein. Dark purple constraints come from a pilot BREAD experiment, dubbed GigaBREAD, which used a custom coaxial horn antenna at the focal point with a low-noise radio-frequency receiver to probe dark photons \cite{BREAD:2023xhc} and axions \cite{GigaBREAD:2025lzq} in the 44--52\,${\rm \mu eV}$ range (10.7--12.5\,GHz).}
    \label{fig:sensitivity}
\end{figure*}

{\textbf{Projected sensitivity.---}}We compute the signal-to-noise ratio (SNR) using
\begin{equation}
{\rm SNR} = \frac{\eta R_{\rm DM}}{\sqrt{\eta R_{\rm DM} + {\rm DCR}}}\sqrt{t_{\rm exp}}\,, \nonumber
\end{equation}
where $\eta$ is the efficiency associated with each of the relevant search modalities discussed above; $R_{\rm DM}$ is the photon production rate from the BREAD setup or the BREAD with additional dielectric layers; ${\rm DCR} \sim 0.01\,{\rm Hz}$ is the dark count rate; and $t_{\rm exp}$ is the averaging time per experiment, defined by a unique set of experimental parameters and configurations.

We assume a total experimental runtime of 1000\,days per decade in mass and adopt an SNR=5 criterion for projected sensitivities, following the conventions of the original BREAD proposal \cite{BREAD:2021tpx}. The results are presented in Fig.\,\ref{fig:sensitivity}. We discuss some features of the projected sensitivities in the following.

\textit{0.1--1\,meV with BREAD:} The search is based on Rydberg–Rydberg transitions in a resonant scheme, with a bin width of $\Delta m_{\rm DM}/m_{\rm DM}=10^{-6}$. As shown in Fig.\,\ref{fig:scanning}, covering the full mass range requires approximately $4\times10^4$ scanning steps when using Rydberg states with $n=31,\cdots,65$. Total number of experiments is therefore $35 \times 4 \times 10^4$, which results in an average measurement time of about $t_{\rm exp}=60$ seconds per experiment. The reported sensitivity lies precisely at the boundary between the background-free and DCR-limited regimes; that is, this sensitivity is achieved in the regime where it is $\propto 1/\sqrt{t_{\rm exp}}$. However, for even longer averaging time the sensitivity scales as $1/t_{\rm exp}^{1/4}$. The efficiency of detecting BREAD photons is $\eta=1$ for this search. 

\textit{0.1--1\,meV with `BREAD+ dielectric layers':} This search is also based on Rydberg–Rydberg transitions, but the photon linewidth is broadened by adding dielectric layers, resulting in bin width of $\Delta m_{\rm DM} / m_{\rm DM} = 1/17$. The entire mass range can be covered with 35 experiments using Rydberg states with $n = 31,\dots,65$, eliminating the need for scanning. The measurement time per experiment is therefore approximately 28 days. In this mass range, the BREAD-produced photon rate is enhanced by a factor of about 300–3000. The detection efficiency for these photons decreases from 1 at the lower end of the mass range to 0.06 at the upper end. The reported sensitivity for this range lies well within the DCR-limited regime, where the sensitivity scales as $1/t_{\rm exp}^{1/4}$ for a longer averaging time.

\textit{1--10\,meV with BREAD:} This search is based on direct ionization of Rydberg atoms. As shown in Fig.\,\ref{fig:scanning}, this mass range can be covered with four experiments using Rydberg states with $n = 55, 75, 100,\,{\rm and}\, 125$. The measurement time per experiment is 250 days. The detection efficiency for these photons decreases from $10^{-4}$ at the lower end of the mass range to $10^{-5}$ at the upper end. The reported sensitivity is DCR-limited, i.e., the sensitivity scales as $1/t_{\rm exp}^{1/4}$ for a longer averaging time.

\textit{1--10\,meV with `BREAD+ dielectric layers':} This search is also based on direct ionization of Rydberg atoms and can similarly be covered with four experiments of $t_{\rm exp}=250\,{\rm days}$. The photon production rate is enhanced by a factor of about $3\times10^3–3\times10^{4}$. The detection efficiency is the same as the previous case without dielectric layers. Note that this differs from the transition-based search, where adding dielectric layers reduces efficiency because the photon linewidth broadens and its frequency shifts off resonance. The reported sensitivity is DCR-limited, i.e., the sensitivity scales as $1/t_{\rm exp}^{1/4}$ for a longer averaging time. The improvement in sensitivity is not as strong as in the lower-mass, transition-based search. This can be understood as follows. In the transition-based case, the improvement comes from two effects: a broader sensitivity bandwidth, which increases the effective integration time per experiment, and an enhancement in photon production. In contrast, the ionization-based search is already broadband in the BREAD setup without dielectric layers. Therefore, the improvement here comes only from the enhanced photon production.

\textit{
\paragraph{Acknowledgments.---} We thank Masha Baryakhtar, Peter Graham, David E. Kaplan, Roni Harnik, Harikrishnan Ramani, Ron Walsworth, and Samuel Wong for useful discussions, and especially Don Fahey for a careful reading of the manuscript and for comments on various experimental aspects.
AB is supported by the National Science Foundation under grant number PHY-2514660 and the Maryland Center for Fundamental Physics.
R.E. is supported by the John Templeton Foundation Award No. 63595
and the University of Delaware Research Foundation
NSF Grant No. PHY-2515007. The work of R.E. was also supported by the Grant 63034 from the John Templeton Foundation and the University of Maryland Quantum Technology Center. S.R. is supported in part by the U.S. National Science Foundation (NSF) under Grant No. PHY-1818899. S.R. is also supported by the DOE under a QuantISED grant for MAGIS. The work of S.R. was also supported by the Simons Investigator Award No. 827042.}

\bibliography{ref}

\clearpage
\widetext
\begin{center}
    \textbf{\Large Supplemental Material}\\ \vspace{5pt}
    \textbf{\large Rydberg Single Photon Detection for Probing 0.1–10 meV Dark Matter with BREAD}\\ \vspace{3pt}
    Abhishek~Banerjee, Reza~Ebadi, and Surjeet~Rajendran
\end{center}
\setcounter{equation}{0}
\setcounter{figure}{0}
\setcounter{table}{0}
\setcounter{page}{1}
\makeatletter
\renewcommand{\theequation}{S\arabic{equation}}
\renewcommand{\thefigure}{S\arabic{figure}}
\renewcommand{\bibnumfmt}[1]{[S#1]}
\renewcommand{\citenumfont}[1]{S#1}

\section{Rate calculations}
In this section, we outline the calculation of a) the transition rate between two rydberg states $i\to f$ and b) the ionization rate from a rydberg state $i$ in the presence of background electric field. 
As the dark matter (DM), which is coherently oscillating converted photon providing the background electric field, it is taken to be classical which oscillating in time with the same frequency as the DM as $\vec E(t)=E_{\rm DM} \cos[\mDM t +\varphi]$ with an amplitude of $E_{\rm DM}$ and a random phase $\varphi$. As the DM velocity
$|v|\simeq 10^{-3}\ll 1$, we omit the space dependent part. 
In the presence of a time dependent electric field, the Hamiltonian becomes time-dependent
as $H_{\rm int} = e \vec E(t)\cdot\vec r$ due to the dipole interaction. 
The problem is similar to calculate the transition due to radiation of frequency $\mDM$. 
\\

\paragraph{\textbf{Transition}.} To calculate the transition probability from $i\to f$ state, we use Fermi's golden rule and obtain
\bea
P_{i\to f}= \left|e \vec E(t)\bra{i}\vec r\ket{f}/2 \right|^2 t^2 \left(\frac{\sin[(\omega_t-\mDM)t/2]}{(\omega_t-\mDM)t/2}\right)^2,
\eea 
where $\omega_t=E_f-E_i$ is the energy difference between two states. 
As, $\sin x/x\to 1$ for $x\lesssim 1$ and $|\sin x/x|\lesssim 1/x^2$ for $x\gtrsim 1$, for $t\lesssim 2/(\omega_t-\mDM)$, we obtain a resonant enhancement. 
Thus for $t\lesssim 2\,{\rm min}[\tau_{\rm coh},\tau_{\rm Ryd}]$, the transition probability can be written as
\bea
P_{i\to f}\approx \left|e E_{\rm DM}\bra{i}\vec r\ket{f}/2 \right|^2 (2\,  {\rm min}[\tau_{\rm coh},\tau_{\rm Ryd}])^2 \,,
\eea
where $\tau_{\rm Ryd}$ is the life time of the rydberg states. 
The transition rate, which is found by averaging the probability over the characteristic time $\Gamma_{i\to f}=\left< P_{i\to f}(t)\right>_\tau/\tau$ can be written as,
\bea
\gamma_{n\to n+i} \approx \frac{2\pi\alpha}{3}|E_{\rm DM}|^2 \, |\left<i|r|f\right>|^2 \times  {\rm min}[\tau_{\rm coh},\tau_{\rm Ryd}]\,.
\eea
For a transition between two rydberg states $n\to n+i$ with $i\ll n$, due to the large radial wave function of the rydberg states  $\left<n|r|n+i\right>\simeq n^2 a_0$ where $a_0=(m_e\alpha)^{-1}$ is the Bohr radius, and $n$ is the principle quantum number~\cite{Gallagher1994}. 
So the rate calculation simplifies to 
\bea
\gamma_{i\to f} \approx \frac{2\pi\alpha}{3}|E_{\rm DM}|^2 \, n^4 a_0^2\,  \times   {\rm min}[\tau_{\rm coh},\tau_{\rm Ryd}]\,.
\eea

In our set up, we are considering a beam of rydberg atom passing through the focusing region of the BREAD experiment which is of the size of $A_{\rm focus}=\pi ( \rm mm)^2$. 
For a beam of velocity $v_{\rm beam}$, traversing the focusing area takes 
$\tau_{\rm sens.}= \sqrt{A_{\rm focus}/\pi}/v_{\rm beam}$
amount of time. 
And if $\tau_{\rm sens.}\lesssim {\rm min}[\tau_{\rm coh},\tau_{\rm Ryd}]$, then each rydberg atom only gets $\tau_{\rm sens.}$ amount of time to interact with the photon. 
Taking this into account, we find
\bea
\gamma_{i\to f} \approx \frac{2\pi\alpha}{3}|E_{\rm DM}|^2 \, n^4 a_0^2\,  \times   {\rm min}[\tau_{\rm coh},\tau_{\rm Ryd},\tau_{\rm sens.}]\,.
\eea
The total photon absorption rate is given by the single-atom rate multiplied by the number of Rydberg atoms $\Gamma_{n\to n+i}=N_{\rm Ryd}\gamma_{n\to n+i}$. 
For a beam of flux $\Phi$, the total number of Rydberg atoms present at any instant in the sensing volume is $N_{\rm Ryd} \simeq \Phi_{\rm beam} A_{\rm focus} \tau_{\rm sens.}$.

One key advantage of a setup such as BREAD is that due to its geometry, the DM induced photon is focused in the focusing area. 
Thus the effective electric field $E_{\rm DM}$ which enters in the rate calculation is the enhanced and can be obtained from the rate calculation of BREAD as $E_{\rm DM}^2=2 R_{\rm DM} m_{\rm DM}/A_{\rm focus}$. 
By plugging in everything, we find the efficiency of photon absorption $\eta = \Gamma_{n\to n+i}/R_{\rm DM}$ as 
\bea
\eta = \frac{4\pi\alpha}{3} \Phi_{\rm beam}  \tau_{\rm sens.} \mDM (n^2 a_0)^2    {\rm min}[\tau_{\rm coh},\tau_{\rm Ryd},\tau_{\rm sens.}].
\eea
An efficiency $\eta\gtrsim 1$ means the rydberg atoms  efficiently absorb all the available photon produced in the experiment. 

\paragraph{\textbf{Ionization}.}
Now we want to calculate the rate for the ionization process due to the background electric field.  
Compare to the previous case,  here we need to calculate the interaction matrix element between one bound state $n$ with energy $E_n$ and one state in the continuum $\ket{\psi_{\rm cont.}}$. 
Since a free state does not have a definite energy, we consider a transition to a  free state with energy in the range $(E_f-\Delta E/2,\,E_f+\Delta E/2)$. 
For the continuum states, we need to calculate the density of states given an energy $E$ which we denote as $\rho(E)$. 

Again using the Fermi's golden rule, we find the transition probability as
\bea
P_{\rm ion}=  \int_{E_f-\Delta E/2}^{E_f+\Delta E/2} && dE  \, \rho(E)\, \left|e \vec E(t)\bra{i}\vec r\ket{\psi_f} \right|^2 (t/2)^2 \times\nonumber\\ 
&& \left(\frac{\sin[(E-E_{n}-\mDM)t/2]}{(E-E_{n}-\mDM)t/2}\right)^2 \,.
\eea
The sinc function is peaked for
$E=E_n+m_{\rm DM}=E_f$, and we get 
\begin{align}
P_{\rm ion}\approx \rho(E_f)\, \left|e \vec E(t)\bra{n}\vec r\ket{\psi_{\rm cont}}\right|^2 \frac{t}{2} 
\int_{-\Delta E t/4}^{\Delta E t/4} dx\, \left(\frac{\sin x}{x}\right)^2\!,\nonumber
\end{align}
where we define $x=(E-E_f)t/2= (E-E_n-\mDM)t/2$. 
The integral evaluates to $\pi$ in the large time limit and like before, by averaging the probability over the characteristic time, we get the ionization rate as,
\bea
\gamma_{n\to {\rm cont}} = \frac{\pi}{2} \times \rho(E_f) \times e^2 |\vec E(t)|^2
 \left|\bra{n\ell m}\vec r\ket{\psi_{\rm cont}} \right|^2\,,
\eea
where we take the bound state has $\ket{i}=\ket{nlm}$ quantum numbers respectively. 
\\

Calculating both $\rho(E_f)$ and $\bra{n \ell m}\vec r\ket{\psi_{\rm cont}}$ involves the normalization of the continuum states. 
Away from the ionization threshold $I_n$, {\it i.e.} for $\mDM\gg |I_n| $, one can approximate the free states as plane waves. 
In that case the absorbed photon energy mostly get converted to the momentum of the free electron, $\vec k$, as $\mDM=|I_n|+|\vec k|^2/(2m_e)$ from the energy conservation. 
Using $\psi_{\rm cont}(\vec r)= 1/\sqrt{L^3} \exp(-i \vec k\cdot\vec r)$ and $\rho(E_f)= (L/2\pi)^3 m_e k\, d\Omega$ (density of states in a given solid angle), we get  
\bea
\gamma_{n\to {\rm cont}} = \frac{\pi}{2} \times \frac{m_e k\, d\Omega}{8\pi^3} \times e^2 |\vec E(t)|^2\times \cos^2\theta
\left| \int_0^{\infty} dr \, r^3 \int_0^{2\pi} d\phi' \int_{-1}^{1} d(\cos\theta') \cos\theta'
e^{-i |\vec k||\vec r| \cos\theta'}
\psi_{n\ell m}(r,\theta',\phi')
\right|^2\,,
\eea
where $\cos\theta$ is the effective angle between the electric field and $\vec k$, and $L$ is the size of the box that is introduced to regulate the integral. Note that, $L$ drops out of the equation and we can safely take the $L\to \infty$ limit to represent the continuum states. 

We can further simplify the above expression if we take the $ns$ state of the rydberg atom as the initial state. 
In that case we have
\bea
\psi_{n00}(r)= \frac{1}{\sqrt{\pi n^5 a_0^3}} \, e^{-r/(na_0)} L^{1}_{n-1}[2r/(na_0)]\,,
\eea
where, $L^{1}_{n-1}[2r/(na_0)]$ is a generalized Laguerre polynomial of degree $n-1$. 
Using this simplification we obtain
\bea
\int_{-1}^{1} d(\cos\theta') \cos\theta'
e^{-i |\vec k||\vec r| \cos\theta'} = \frac{2 i }{k^2 r^2} \times \left[-k r \cos(k r) + \sin (k r)\right]\,,  
\eea
and plugging the above expression we further get,
\bea
\gamma_{n\to {\rm cont}} = \frac{m_e e^2 |E(t)|^2}{\pi n^5 a_0^3}\times  \cos^2\theta d\Omega \times \frac{1}{k^3}  \times 
\left| \int_0^{\infty} dr \, r \left[-k r \cos(k r) + \sin (k r)\right] e^{-r/(na_0)} L^{1}_{n-1}[2r/(na_0)]
\right|^2\,.
\eea
By using the fact that $\cos^2\theta d\Omega=4\pi/3$, and definign $\tilde{r}= r/(n a_0)$ and $\tilde{k}=k na_0$, we can further simplify the above expression as
\bea\label{eq:gamma_ion_whole}
\gamma_{n\to {\rm cont}} = \frac{16\pi  |E(t)|^2 \, a_0}{3} \times \frac{(n a_0)^2}{(kn a_0)^3}  \times 
\left| \int_0^{\infty} d\tilde{r} \, \tilde{r} \left[-\tilde{k} \tilde{r} \cos(\tilde{k} \tilde{r}) + \sin (\tilde{k} \tilde{r})\right] e^{-\tilde{r}} L^{1}_{n-1}(2\tilde{r})
\right|^2\,.
\eea
Note that the integral is given in terms of  dimensionless quantities. 
To evaluating the above integral for a generic $n$, we find that integral is proportional to 
\bea
\int_0^{\infty} d\tilde{r} \, \tilde{r} \left[-\tilde{k} \tilde{r} \cos(\tilde{k} \tilde{r}) + \sin (\tilde{k} \tilde{r})\right] e^{-\tilde{r}} L^{1}_{n-1}(2\tilde{r}) \propto \frac{8 \, n \,\tilde k^3}{(1+\tilde k^2)^{2+n}}= \frac{8 n (ka_0 n)^3}{(1+ k^2a_0^2 n^2)^{2+n}}\,,
\eea
with the proportionality constant fixed to $n 
(k^2a_0^2n^2)^{n-1} $ for large $k$, {\it i.e.} for $ka_0 n\gg 1$ and to $(-1)^{n-1} n (n^2+2)/3$ for $k\to 0$. 
However, very close to the threshold, the electron wavefunction can not be approximated as plane-waves as one expects corrections to this coming from the columbic part of the potential. 
In that case, the continuum wave function of the Columb potential should be used as discussed in~\cite{Ovsiannikov2011,Glukhov_2010}. 
Thus, in what follows we will only consider the case of $ka_0n\gg 1$. 
By solving Eq.~\eqref{eq:gamma_ion_whole} for some values of $n$, we find the above simplification is a conservative estimate and the actual ionization cross-section is much larger than our approximated value close to the threshold. 
\\

Plugging everything together, we obtain the ionization rate as
\begin{align}\label{eq:gamma_ion_tot}
\gamma_{n\to {\rm cont}} &= \frac{1024\pi}{3} E_{\rm DM}^2 a^3_0 n^4 \frac{(k na_0)^3}{(1+ k^2a_0^2n^2)^{4+2n}} \times n^2  (k^2a_0^2n^2)^{2n-2} \nonumber\\
&\approx \frac{1024\pi}{3} E_{\rm DM}^2 a^3_0 n^4 \left(\frac{|I_n|}{m_{\rm DM}}\right)^{4+2n} \left(\frac{m_{\rm DM}}{|I_n|}-1\right)^{3/2} \times n^2
 \left(\frac{m_{\rm DM}}{|I_n|}\right)^{2n-2}\,,
\end{align}
where we have used $k^2/2m_e= \mDM-|I_n|$ with $|I_n|=m_e\alpha^2/(2n^2)$ and have approximated $(k^2a_0^2n^2)^{2n-2}\simeq (\mDM/|I_n|)^{2n-2}$.  
As expected we find that the Ionization rate is small close to the threshold and reaches a maximum value for $\mDM\sim \mathcal{O}(|I_n|)$ before falling off as a polynomial in incident photon energy away from the threshold. 
By expressing the ionization rate in term of cross-section $(\sigma)$ with $\gamma_{n\to {\rm cont}}= (E_{\rm DM}^2/2\mDM) \times \sigma$, we find
\bea
\sigma &\propto & a_0^3 n^6 \mDM     \left(\frac{|I_n|}{m_{\rm DM}}\right)^{4+2n} \left(\frac{m_{\rm DM}}{|I_n|}-1\right)^{3/2}\times  \left(\frac{m_{\rm DM}}{|I_n|}\right)^{2n-2} \nonumber\\
&\propto& a_0^3 n^6 \mDM     \left(\frac{|I_n|}{m_{\rm DM}}\right)^{6} \left(\frac{m_{\rm DM}}{|I_n|}-1\right)^{3/2}\nonumber\\
&\propto& \frac{a_0 n^4}{m_e}   \left(\frac{|I_n|}{m_{\rm DM}}\right)^{5} \left(\frac{m_{\rm DM}}{|I_n|}-1\right)^{3/2} \sim \frac{1}{n^3}\frac{e^2}{m_e} \frac{(m_e\alpha^2/2)^{5/2}}{\mDM^{7/2}}\,,
\eea
in the large momentum limit {\it i.e.} for $nka_0\gg 1\Rightarrow \mDM\gg |I_n|$. 
Thus away from the threshold we obtain the result given in~\cite{BetheSalpeter1957}. 
Even for $\ell=1$ states, we obtain the same result as~\cite{BetheSalpeter1957} in the large-momentum (away from the threshold).   

As discussed previously, $n^2 (\mDM/|I_n|)^{(2n-2)}$ term in the rate calculation is valid away from threshold. It turns out that extending this estimate to the threshold region overestimates the rate \cite{Ovsiannikov2011}. We conservatively drop $n^2$ and use 
$\gamma_{n \rightarrow {\rm cont.}} \simeq (1024\pi/3)E_{\rm DM}^2 n^4 a^3_0\,(I_n/m_{\rm DM})^6 \left(m_{\rm DM}/I_n-1\right)^{3/2}$ in the main text to obtain the experimental reach. 
Note that close to the threshold, the rate obtained from our estimate is consistent with the numerical results as well as the $n$-scalings in Refs.~\cite{Ovsiannikov2011,Glukhov_2010}. 
We leave a detailed computation of near-threshold ionization rate for future work. Close to the threshold, the plane wave approximation of the electron wave-function breaks down as the corrections due to the columbic potential start to become important. In this case one should use the continuum (positive-energy) Coulomb wavefunctions for the final electron and evaluate the dipole matrix element between the bound Rydberg state and those continuum states.

We computed the rates for $\ell=0$, $m=0$ states. These results agree with the result of~\cite{BetheSalpeter1957} for $\ell=0$ states. For low-$\ell$ states, the power-law fall-off away from threshold has only a mild dependence on the $\ell$~\cite{BetheSalpeter1957,Ovsiannikov2011}. Also, the threshold ionization rate and its $n$ scaling does not depend on $\ell$ at the leading order~\cite{Ovsiannikov2011}. Thus, we justify the use of the conservative order-of-magnitude estimate mentioned above.

\end{document}